\documentclass{iau}
\usepackage{graphicx}

\newcommand\lsim{\mathrel{\rlap{\lower4pt\hbox{\hskip1pt$\sim$}}
        \raise1pt\hbox{$<$}}}
\newcommand\gsim{\mathrel{\rlap{\lower4pt\hbox{\hskip1pt$\sim$}}
        \raise1pt\hbox{$>$}}}
\newcommand\propsim{\mathrel{\rlap{\lower4pt\hbox{\hskip1pt$\sim$}}
        \raise1pt\hbox{$\propto$}}}

\newcommand{\hr}{\,\mathrm{hr}}

\newcommand{\Hz}{\,\mathrm{Hz}}

\newcommand{\Msun}{\mathrm{M}_{\odot}}

\newcommand{\pc}{\mathrm{pc}}

\newcommand{\BH}{\star}

\newcommand{\SMBH}{\bullet}

\title[Detecting gravitational waves from the galactic center with pulsar timing] 
{Detecting gravitational waves from the galactic center with Pulsar Timing} 

\author[A. Ray, B. Kocsis \& S. Portegies Zwart]   
{Alak Ray$^1$
Bence Kocsis$^2$
 \and Simon Portegies Zwart$^3$}

\affiliation{$^1$Tata Institute of Fundamental Research, Mumbai 400005,
India 
\\[\affilskip]
$^2$Institute for Advanced Study, Einstein Drive, Princeton, NJ 08540,
USA 
\\[\affilskip]
$^3$Sterrewacht Leiden, Leiden University, P.O. Box 9513, 2300 RA Leiden, 
The Netherlands
\\[\affilskip]
}

\pubyear{2013}
\volume{303}  
\pagerange{119--124}
\jname{The Galactic Center}
\editors{Lorant Sjouwerman, Jurgen Ott \& Cornelia Lang, eds.}
\begin{document}

\maketitle

\begin{abstract}
Black holes orbiting the Super Massive Black Hole (SMBH) Sgr A* in the Milky-way 
galaxy center (GC) generate gravitational waves. The spectrum, due to stars and black holes, 
is continuous below 40 nHz while individual BHs within about 200 AU of the central SMBH 
stick out in the spectrum at higher frequencies. The GWs can be detected by timing radio 
pulsars within a few parsecs of this region. Future observations with the Square 
Kilometer Array of such pulsars with sufficient timing accuracy
may be sensitive to
signals from intermediate mass BHs (IMBH) in a 3 year observation baseline. 
The recent detection of radio pulsations from the magnetar SGR J1745-29 very near the 
GC opens up the possibilities of detecting millisecond pulsars (which can be used
as probes of the GWs) through lines of sight with 
only moderate pulse and angular broadening due to scattering. 
\end{abstract}

\firstsection 
\section*{Introduction}
The central region of the galaxy has a dense population of visible stars and 
most likely, also of compact objects, such as black holes, neutron stars and Intermediate Mass Black
Holes (IMBH). 
In combination with the supermassive blackhole (SMBH) Sgr A*, these BHs or IMBH
can make their presence felt e.g., by gravitational waves (GW).
The Laser Interferometric Gravitational Observatory will attempt to
detect GW by change of baselines (the ``strains" or the fractional changes) 
between two arms of an interferometer.
%
The alternate idea of detecting GWs by monitoring the arrival times of radio pulses from 
neutron stars (pulsar timing) was proposed by \cite{Saz78} and \cite{Det79}.
The formulation in terms of an array of pulsars and their correlations of pulse
arrival time residuals was given by \cite{Hel83}. 
While LIGO and Advanced LIGO are sensitive to a frequency range 10Hz -100 kHz, the Pulsar Timing
Arrays (PTAs) probe
GWs typically in the 300 picoHz to 100 nanoHz.
In this frequency band, the strongest gravitational waves are likely to be from binary systems
in which two massive black holes orbit one another resulting from galaxies merging with one another.
For the GW source in the central part of our own galaxy, the prospects of resolving
individual objects through GW measurements improve closer to Sgr A* even if the
number density of objects steeply increases inward, quite unlike the case of imaging
by electromagnetic techniques.
In this paper we discuss future prospects of detecting gravitational waves from the galactic center by
accurate timing of pulsars which may be discovered in the galactic center with 
sensitive telescope arrays in the future (we refer to \cite{Koc12} for more details).
%

\section*{Sources of gravitational waves in the galactic center}

The central region of the galaxy, very likely, has a dense population of compact objects, 
including about 20,000 stellar mass black holes (BHs; \cite{Mor93}; \cite{Fre06a} 
and perhaps a few IMBH of mass $10^3 M_{\odot}$ 
IMBHs 
(\cite{Por06}) 
may be created by the collapse of Pop III stars in the early universe
(\cite{Mad01}),
runaway collisions of stars in the cores of globular clusters (\cite{Por02}, \cite{Fre06b}),
or the merger of stellar mass black holes (\cite{OLe06}, \cite{Gul04}). 
The globular clusters can be tidally stripped when they sink to the galactic nucleus as a result of dynamical
friction, leaving their IMBHs behind in the galactic nucleus.
\cite{Por06} predict that the inner $10\,$pc
of the GC hosts 50 IMBHs of mass $M\sim 10^{3}\Msun$.
These objects as also the stellar mass BHs are much more massive than regular stars 
populating the GC and they segregate and settle to the core of the central star cluster. 
For a circular binary of an object ($m_{\BH}$) e.g. an IMBH or a BH, orbiting around a SMBH of mass 
$M_{\SMBH}$ the orientation averaged RMS strain generated
at distance $D$ from the source in one GW cycle is
$$h_0(f) = \sqrt{\frac{32}{5}} \,\frac{M_{\SMBH}m_{\BH}}{D\, r(f)}
  = 8.8\times 10^{-15} m_{3} D_{\rm pc}^{-1} f_{8}^{2/3}\,,$$
where $D_{\pc}=D/{\pc}$, $m_{3}=m_{\BH}/(10^3 \Msun)$, $f_{8}=f/(10^{-8}\Hz)$ 
with the GW frequency $f = 2 f_{orb}$, $r(f)$ is the orbital radius around SMBH
and the index 0 on h stands for zero eccentricity.
The total GW signal with frequency $f$ (the ``characteristic spectral amplitude"
$h_c^2$) from a population of sources (with a number in a shell
$dN = 4 \pi r^2 n_{\BH}(r) dr$ where $n_{\BH}(r)$ is the number density of objects) 
is:
$$h_c^2(f) =  \frac{8\pi}{3} r^3 n_{\BH}\hspace{-1pt}(r)\, h_0^2 =
\frac{256\pi}{15} \frac{M_{\SMBH} m_{\BH}^2}{D^2} \, (\pi M_{\SMBH} f)^{-2/3} \, n_{\BH}\hspace{-1pt}[r(f)]\,.$$
The GW signal generated by a population of objects (the ``foreground") is smooth if the average number 
per $\Delta f$
frequency bin satisfies $\langle \Delta N\rangle \gg 1$. 
The GW spectrum becomes spiky
(with $\langle \Delta N\rangle \leq 1$) above
a critical frequency $f_{\rm res}$ that depends on the number of objects within 1pc of the GC and
on the timing observation span.
Sources within $r_{\rm res}$ generate distinct spectral peaks above frequency $f_{\rm res}$.
These sources are {\it resolvable}.
%
The GW spectrum transitions
from continuous to discrete at higher frequencies inside the PTA band
and 
there may be a number of resolvable sources (\cite{Koc12}).

\begin{figure}[b]
\vspace*{-0.2 cm}
\begin{center}
 \includegraphics[width=4.2in]{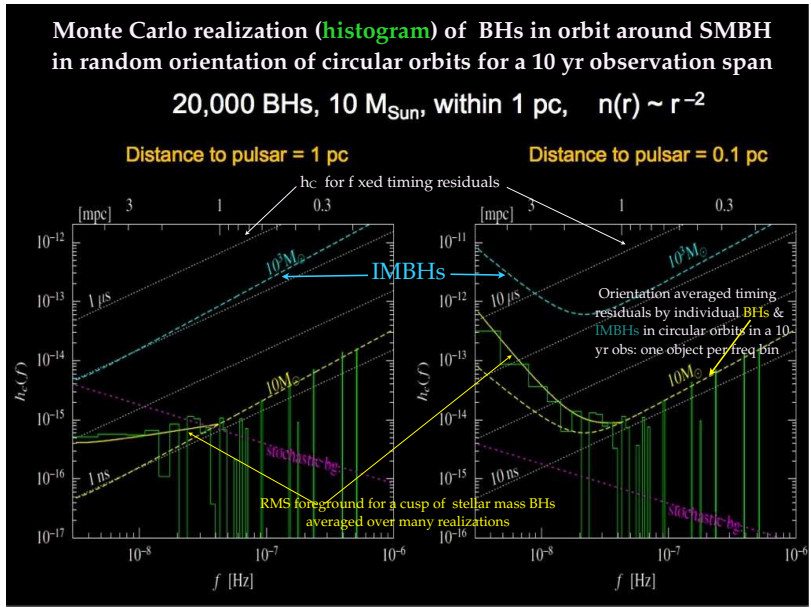} 
 \caption{Phase space of characteristic strain amplitude $h_c$ vs. GW frequency $f$
for detection by a pulsar at 1\,pc (left) and 0.1\,pc (right) from the Galactic Center.
Dotted white lines show the orientation-averaged $h_c$ for fixed timing residuals.
Yellow and cyan dashed lines show respectively the orientation-averaged timing residuals
caused by individual stellar BHs and IMBHs on circular orbits
in a 10 yr observation, assuming 1 object per frequency bin.
Green lines (histogram) show the timing residuals for a random realization of $10\Msun$
stellar BHs in the cluster (20,000 BHs within 1 pc with number density $\propto r^{-2}$).
At high $f$, only few bins are occupied, generating a spiky signal.
Magenta dashed lines show the cosmological RMS stochastic GW background.
}
   \label{fig1}
\end{center}
\end{figure}

\section*{Using Pulsar Timing to detect GWs}
If pulsars are observed repeatedly in time 
for an observation span $T = 10$ yr and with $\Delta t = 1$  week, 
the range of GW frequencies is:
$3 \times 10^{-9} \; \rm Hz (3 nHz) < f < 3 \times  10^{-6} \; \rm Hz (3000 nHz)$.
The cosmological GW background from the whole population of MBHBs
is actually an astrophysical "noise" for the
purpose of measuring the GWs of objects orbiting SgrA*.
The characteristic GW amplitudes (either of a stochastic background or of a resolvable source) 
can be translated into into a ``characteristic timing residual" $\delta t_c (f)$ 
corresponding to
a delay in the time of arrivals of pulses due to GWs, after averaging over the sky position
and polarizations. 
The timing signal to noise ratio (S/N) is proportional to $h_c/(2\pi f)$ which incorporates the
$(fT)^{1/2}$ factor that accounts for the residual built-up over the number of GW cycles, 
$h_c^2$ being the characteristic spectral amplitude.
The distance within which a PTA could measure the GWs of an individual source
with a fixed timing precision $\delta t=10 \, \delta t_{10}\, \ns$ is
$D_{\delta t} = 14\, m_3 \delta t_{10}^{-1} T_{10}^{1/2} f_8^{1/6} \,\pc$.
The expressions for the strain amplitudes (see Eq. (13) and (16) of \cite{Koc12}) 
for a resolved binary source and that for the
stochastic GW background (dominated by SMBHB inspirals) show that
the GWs from an individual BH in the GCs rises above the stochastic GW background within a distance
$D_{\rm bg} = 8.7\, m_3 T_{10}^{1/2} f_{8}^{11/6}  \,\pc\,$.
Thus a pulsar within $D_{\delta t}$ and $D_{\rm bg}$ to the GC could be used to detect GWs from
individual objects in the GC. 
%
%
Prospects of GW detection from the GC depends upon future
discoveries of pulsars near the GC and timing them with sufficient accuracy.
Active radio pulsars/ ms PSRs may segregate to the outer parts of the GC as heavier objects sink inward
(\cite{Cha02}).
\cite{Liu12} examined the expected timing accuracy of pulsars in the GC
and found that the $1\,\hr$ timing accuracy of SKA is expected to be between $10$--$100\,\mu{\rm s}$
for regular 
pulsars. Our results summarized in Fig. 1 indicate that the necessary accuracy to detect timing 
variations associated to individual
$10\,\Msun$ BHs within 1\,mpc requires much higher timing accuracy, 
which might be prohibitively difficult
even with MSPs with a factor of 100--1000 better timing accuracy. 
However, the net variations caused by a population of these objects is
detectable between 2--5 mpc at these accuracy levels. As seen from Fig 1, a
$10$--$100\,\mu{\rm s}$ timing accuracy 
is sufficient to individually resolve or rule out the existence of $10^3\,\Msun$ IMBHs 
within 5 mpc from SgrA*.
As a point of reference, the Parkes Pulsar Timing Array (PPTA) project aims to time 20 
pulsars, with an rms of 100 ns 
over five years 
(\cite{Wen11}).

\section*{Pulsar discovery near the Galactic Center}

A large population of pulsars may reside inside the GC (\cite{Pfa04}, \cite{Lor04}). 
\cite{Wha12}
predict as many as 100 canonical PSRs and a  
larger population of ms PSRs in the central parsec of the GC. 
The discovery of GC magnetar SGR J1745-29 with NuSTAR 
and subsequently 
in the $1.2- 18.95$ GHz radio bands
(\cite{Bow14}, \cite{Spi14})
shows that the source angular sizes are consistent with scatter 
broadened size of SgrA* at each radio frequency. This demonstrates that the two sources, separated by 3" 
(0.12 pc in projection) are both located behind the same hyperstrong scattering medium, e.g. a ``thin" screen 
at ~6 kpc from the GC. The pulse broadening timescale at 1 GHz 
(\cite{Spi14}) is several orders of magnitude lower than the scattering predicted by NE2001 model. 
The scattering in the GC is 
lower than previously thought. 
The scattering material could be patchy at small angular scales
(\cite{Spi14}) and it may be 
possible to peer through such 
``keyholes" in the radio bands for a region around GC.
\cite{Che13} using Monte Carlo simulation 
estimate an upper limit of ~950 potentially observable radio loud pulsars in GC.
However,
\cite{Dex13} point out that despite several 
deep radio surveys, no ordinary pulsars have been detected 
very close to the GC
and suggest an intrinsic deficit in the ordinary pulsar population. 
This analysis does not constrain the millisecond pulsar population significantly as yet.
Since it now appears that they can be observed at somewhat lower frequencies because of the nature of the
scattering, deeper surveys at $\geq 8 $ GHz may be able to discover them. 
Future discovery of such pulsars in the GC may facilitate the long term timing and search for
gravitational waves due to IMBHs.

BHs in orbit around SMBH SgrA* generates a continuous GW spectrum with $f < 40$ nHz. 
Individual BHs within 1 milliparsec to SgrA* stick out in the spectrum at higher frequency.
GWs can be resolved by timing PSRs located within this region. A 100 ns - 10 $\mu$s timing 
accuracy with SKA will be sufficient to detect IMBHs ($10^3 \; M_{\odot}$), if they exist, in a 3 yr observation
if the PSRs are $0.1- 1$ pc away from SgrA*. 
Unlike electromagnetic imaging, resolving individual binaries via GWs detected by pulsar timing 
will improve as one probes the region closer to SgrA*. 
 
%

%
%
%



\end{document}